\documentclass[aps,prl,twocolumn,showpacs,groupaddress,
superscriptaddress
]{revtex4-1}
\usepackage{amssymb}
\usepackage{amsmath}
\usepackage{graphicx}
\usepackage{color}

\newcommand{\nnabla}{\mathbf \nabla}

\newcommand{\req}[1]{Eq.~(\ref{#1})}
\newcommand{\reqs}[1]{Eqs.~(\ref{#1})}
\newcommand{\rref}[1]{(\ref{#1})}
\renewcommand{\a}{\mathbf{a}}

\renewcommand{\b}{\mathbf{b}}

\newcommand{\n}{\mathbf{n}} 
\renewcommand{\r}{\mathbf{r}}
\newcommand{\R}{\mathbf{R}}
\renewcommand{\k}{\mathbf{k}}

\newcommand{\beq}{\begin{equation}}
\newcommand{\eeq}{\end{equation}}
\newcommand{\be}{\begin{equation}}
\newcommand{\ee}{\end{equation}}
\newcommand{\beqa}{\begin{eqnarray}}
\newcommand{\eeqa}{\end{eqnarray}}
\newcommand{\bea}{\begin{eqnarray}}
\newcommand{\eea}{\end{eqnarray}}
\usepackage{amssymb}
\usepackage{array}
\usepackage{amsmath}
\usepackage{graphicx}\usepackage{graphics}
\usepackage{dcolumn}
\usepackage{bm}

\newcommand{\hsigma}[2]{\hat{\sigma}_{#1}^{#2}}

\newcommand{\hb}{\hat{b}}
\newcommand{\hLambda}{\hat{\Lambda}}

\usepackage{amsfonts}

\begin{document}

\title{Localization and critical diffusion of quantum dipoles in two
dimensions.}
\author{I.L. Aleiner}
\author{B.L. Altshuler}
\affiliation{Physics Department, Columbia University, New York, N.Y. 10027, USA}
\author{K. B. Efetov}
\affiliation{Theoretische Physik III, Ruhr-Universit\"{a}t Bochum, 44780 Bochum, Germany}
\affiliation{International Institute of Physics, UFRN, 59.078-400 - Natal/RN- Brazil}
\date{\today}

\begin{abstract}

We discuss quantum propagation of dipole excitations in two
dimensions. This problem differs from the conventional Anderson
localization due to existence of long range hops. We found that the
critical wavefunctions of the dipoles always exist which
manifest themselves by a scale independent diffusion constant.
If the system is T-invariant the states are critical
for all values of the parameters. Otherwise, there
can be a ``metal-insulator'' transition between this ``ordinary'' diffusion
and the Levy-flights (the  diffusion
constant logarithmically increasing with the scale). These results
follow from the two-loop analysis of the modified non-linear
supermatrix $\sigma$-model.

\end{abstract}

\maketitle

  Anderson \cite{Anderson} showed that a quenched
disorder can localize a quantum particle, {\em i.e.}
completely suppress its diffusion. Later \cite{aalr,GLK}, it was realized
that in  two dimensions ($2D$) localization occurs for an
arbitrary weak disorder.
This conclusion was reached by studying
the scaling behavior of the dimensionless Thouless conductance
$g(L)$ as a function of the linear size of the system $L$ (the
observable electrical conductance of the system of the $e$ charged
particles is given by $g\times e^2/\hbar$, and we will set Planck
constant $\hbar=1$ hereinafter). Localization implies that $g_{L}\to 0$ as $L \to \infty$.
This is always true when the time reflection symmetry (T-invariance)
is broken (so called unitary ensemble, GUE).
For T-invariant systems it is still correct
if the orbital and spin degrees of freedom
are decoupled or when the particles have an integer spin (orthogonal
ensemble, GOE). For the particles with half-integer spin,
the theory \cite{hln} predicts that the spin-orbital coupling causes
antilocalization $g(L \to \infty) \to \infty$
if disorder is weak,
while for a stronger disorder $g(L \to \infty) \to 0$ (metal-insulator
transition for symplectic ensemble)
\cite{footnote0}.

Besides the current carrying charged particles  important objects in
the many-body theory are neutral excitations (NEX)
-- the bound states of two particles with opposite charges.
One can name excitons in semiconductors, optical phonons
in polar crystals,
dipole excitations in granular superconductors, vacancy-interstitial
excitations in  Wigner crystals or vortex lattices, etc. Being the
lowest energy excitations, the NEX not only
determine the low temperature energy transport but also
provide a thermal bath for more energetic
charge excitations, thus influencing the charge transport.
As for any neutral particles, the number of NEX is not
conserved ({\em e.g.} electron and hole can
annihilate each other). Each NEX has a finite energy and 
cannot simply disappear. However, their number
non-conservation
facilitates long range hops mediated by virtual
photons (this leads also to dipole-dipole interactions).
In this Letter we investigate the effect of the 
long-range hops on the localization of NEX.

We will be interested in 2D quantum
dipoles -- NEX whose
annihilation [creation] operators $\hb_\alpha(\r)\
[\hb_\alpha^\dagger(\r)]$ are characterized by the additional
index $\alpha=x,z$. The pair $(\hb_x, \hb_z)$ transforms
under rotations similar to a vector in 2D plane
$(x,z)$. For a small density of NEX, we can
neglect the interaction between them and use a bilinear form of
$\hb_\alpha,\hb_\alpha^\dagger$ as the Hamiltonian
$\hat{\cal H}=\int d^2\r_1d^2\r_2
\hb_\alpha^\dagger(\r_1)H_{\alpha\beta}(\r_1-\r_2)\hb_\beta(\r_2)$,
(we imply summation over the repeated
indices $\alpha,\beta=x,z$ hereinafter).
The absorption-emission of the virtual photons
(with infinite speed) results
in a long-range hopping term \cite{footnote1}
\be
H^{lr}_{\alpha\beta}={\lambda 
\left(\delta_{\alpha\beta}|\r|^2-2r_{\alpha}r_{\beta}\right)}/{(2\pi |\r|^4)},
\label{dipole-hop}
\ee
additional to the local Hamiltonian
\begin{subequations}
\label{H}
\be
H^{sr}_{\alpha\beta}=
\delta(\r_1-\r_2)\left[H^0_{\alpha\beta}(-i\nnabla)+V_{\alpha\beta}(\r)\right].
\ee
Constant $\lambda$ in \req{dipole-hop} encodes dipole-photon 
transitions  matrix
elements and the energy denominators.

The hops \rref{dipole-hop} qualitatively modify the Anderson localization.
We derived and analyzed the renormalization group (RG)
equations describing the scaling of the
Thouless conductance $g(L,\lambda)$.
For the GOE we found the {\em stable}
fixed line $g_c(\lambda)$ on the $g - \lambda$ plane, which corresponds to the
the critical state, see Fig.~\ref{fig1}a.
Remarkably, $g_c(\lambda)$ may be large, which
makes this line accessible in the perturbative RG. For the
GUE, we discover the {\em unstable}
line $g_{\rm mit}(\lambda)$ separating the antilocalization
and the localization behaviors, see Fig.~\ref{fig1}b.
We argue that the latter is the precursor
of the critical state similar to the one in
the GOE, which, unfortunately,
 occurs at $g \lesssim 1$ and thus, is out of
reach of the perturbative RG analysis.

\begin{figure}
\includegraphics[width=0.86\columnwidth]{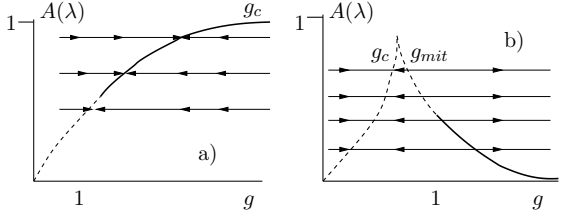}
\caption{RG flows for (a) GOE and (b)
  GUE, and $A(\lambda)<1$ is defined in \req{A}. Solid lines
  are accessible by the perturbative RG, and dashed ones are conjectures.}
\label{fig1}
\end{figure}

The quantum motion  of the dipoles in the rotationally symmetric clean system is described by
\be
{H}^0_{\alpha\beta}
= \frac{\k^2\delta_{\alpha\beta}}{2m_1(\k)}+\frac{k_{\alpha}k_{\beta}}{2m_2(\k)}
+ h\left[\hsigma{y}{}\right]_{\alpha\beta},
\label{a7}
\ee
where $m_{1,2}\left(\k \right)$ are analytic functions of the
momentum $\k$, invariant with respect to the lattice  symmetry group.
The Pauli matrices $\hsigma{i}{},\ i=x,y,z$ act in $2\times
2$ space of dipole components $xz$.
The last term in \req{a7} breaks the T-invariance by the magnetic field $\propto h$
(The Hamiltonian of NEX cannot contain the vector potential).
At $h=0$, we can diagonalize ${H}^0+H^{lr}$ (see Fig.~\ref{fig2}a):
\be
E_-(\k) =  \frac{\k^2}{2m_1(\k)}; \ E_+(\k) =E_-(\k)+ \frac{\k^2}{2m_2(\k)} + \lambda.
\label{clean}
\ee
Note that $E_\pm(\k)$ are analytic functions of $\k$,  even though
$H^{lr}$ lifts the degeneracy  at $k=0$  protected by symmetry.

The $2\times 2$ matrix $\hat{V}(\r)$ in $xz$-space is a Gaussian
disorder breaking all the system spatial symmetries,
\begin{equation}
\left\langle \hat{V} (\r) \otimes \hat{V}(\r^\prime) \right\rangle
  =\delta (\r-\r^\prime)
\sum_{i=0,x,y,z}u_{i}^2\hsigma{i}{} \otimes \hsigma{i}{},  \label{a9}
\end{equation}%
where $\hsigma{0}{}\equiv \hat{\openone}$.
For T-invariant systems, $u_{y}=0.$ The rotational
symmetry after the disorder averaging requires $u_{x}=u_{z}$.
The elastic mean free time
 $\tau$ is given by $1/\tau=\pi\nu
 \sum_{i=0,\dots,z} u_{i}^2$, with the total density of states being
\end{subequations}
\begin{equation}
\nu\left( \epsilon \right)=\sum_{\pm} \int \delta \left[ \epsilon - E_\pm (\k)\right] {d^2{\k}}/{\left( 2\pi \right) ^{2}}.
\label{a10}
\end{equation}%
We assume that the disorder is weak, $\epsilon \tau \gg 1$.

For the specific form of \req{dipole-hop},  the
time evolution
$\hat{U}(t)=\exp\left(it\int d^2\r_1d^2\r_2
\hb_\alpha^\dagger(\r_1)H_{\alpha\beta}^{lr}(\r_{12})\hb_\beta(\r_2)\right)$
can be described locally by introducing
additional fields
\be
\begin{split}
&\hat{U}^{lr}(t)=\int {\cal D}\a (\r){\cal D}\a^*(\r)
\exp\left(\frac{i\lambda t}{2}\int d^2 \r \hat{c}^\dagger\hat{M}\hat{c}\right);
\\
& \hat{c}^\dagger=
\left[\left(a^*_x,a^*_z\right);
  \left(b^\dagger_x,b^\dagger_z\right)\right]^{ab};\
\hat{M}=\begin{pmatrix} \nnabla^2 \hsigma{z}{}; & - \nabla_\alpha \hsigma{\alpha}{}  \\
 \nabla_\alpha \hsigma{\alpha}{};
& \openone\end{pmatrix}^{ab}
\end{split}\raisetag{50pt}
\label{local}
\ee
Due to the locality of the Poisson equation describing virtual photons, operator $\hat{M}$
is differential rather than
integral one and this enables us to develop a
renormalizable theory of localization.

To understand the effect of the long hops
\rref{dipole-hop} on the localization, consider the dipole
with  high energy, $\epsilon \gg \lambda$, see Fig.~\ref{fig2}a. For
the clean system, the wave functions are comprised by
plane waves with wavevectors $k_\pm(\epsilon)$ and
oscillate rapidly. The long-range hops are thus irrelevant.

\begin{figure}[h]
\includegraphics[width=\columnwidth]{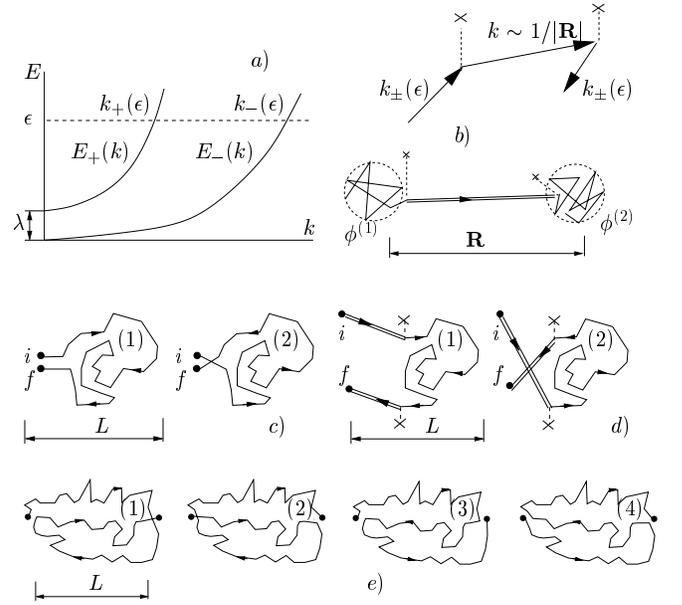}
\caption{a) Band structure for the clean system; b) Long hop assisted
  by the impurities (crosses); c) Semiclassical paths for one-loop
  weak localization; d) Long range hops and semiclassical paths
  for two-loop renormalization; e) Two loop
  semiclassical paths including only local evolution.}
\label{fig2}
\end{figure}

\begin{subequations}
\label{evolution}
In the presence of disorder a dipole is characterized by the
probability density $\n(r)$ of finding it at a point  $\r$. Under  the
assumption of  Markovian time evolution,
\be
\partial_t n(\r)=\int d^2 \R\, W(\R)\, \left[n(\r+\R)-n(\r)\right].
\label{evolutiona}
\ee
At large enough distances and times, \req{evolutiona} reduces to the diffusion equation with diffusion constant $D$ determined by the rates kernel $W(\R)$:
\be
D = \int  d^2 \R\, W(\R)\, (\R^2/4).
\label{evolutionb}
\ee
\end{subequations}

To estimate $W(\R)$, consider the transition between wave-packets $\phi_\alpha^{(1,2)}$,
formed by plane-waves close to the energy shell and normalized  $\int d^2\r
\phi^{(i)}_\alpha [\phi^{(i)}_\alpha]^*=1$, see
Fig.~\ref{fig2}b.
 The disorder can {\em e.g.} scatter   $\phi_\alpha^{(1)}$
into a virtual small momentum state, this state can be transferred far by the
Hamiltonian \rref{dipole-hop}, and then scattered by disorder back onto the
energy shell  (upper part of Fig.~\ref{fig2}b). The amplitude for
this process, $J(\R)$, is  given by
\be
{J}  \simeq \frac{\lambda\left(\R^2 {v^{(1)}_\alpha (v^{(2)}_\alpha)^*
-2{v^{(1)}_\alpha}}
    R_\alpha R_\beta (v^{(2)}_\beta)^* \right)}{2\pi \epsilon^2 |\R|^4},
\label{amplitude}
\ee
where $v^{(j)}_\alpha  = \int d^2\r  \phi_{\alpha^\prime}^{(j)}V^{\alpha^\prime\alpha}$.
We substitute \req{amplitude} into the golden rule formula, average
it over the disorder \rref{a9}, and use the normalization of $\phi_{\alpha}^{(j)}$.
We find
\be
W_{lr}(\R)=2\pi\nu \langle |J(\R)|^2 \rangle =
  \lambda^2/\left(4 {\pi^3\nu \epsilon^4 \tau^2 |R|^4}\right),
\label{W}
\ee
Substituting \req{W} into \req{evolutionb} and taking into account
only hops with $L<|R|<L + \delta L$,
we obtain a positive correction to the Thouless conductance
$g=\nu D$:
\be
\delta g_{lr} = \delta {\ell} {A}/{(2\pi^2)} ; \quad \ell=\ln({L}/{L_0}),
\label{g1}
\ee
where $L_0$ is a microscopic scale, and [see \req{W}]
$
 A= {\lambda^2}/\left(4{\tau^2\epsilon^4}\right);\ \tau\epsilon\gg 1,\
 \epsilon \gg \lambda.
 $
For $\epsilon$ 
in resonance with $E_\pm$, the summation of all orders of perturbation
theory in $\lambda$ gives
 the Breit-Wigner type formula
 \be
 A= \frac{4\lambda^2\tau^2}{\left[4+\tau^2\xi_+^2\right]
\left[4+\tau^2\xi_-^2\right]} < 1,\ \xi_\pm\equiv \epsilon -E_{\pm}(0),
 \label{A}
 \ee
{\em i.e.} even at  $\lambda > 1$ ($\epsilon \to \lambda$), the contribution
of the long-range hops cannot exceed the unitary limit.

Levy flight term \rref{g1} is not the only  logaritmical contribution to the
conductance.  The interference of close
time-reversed paths ($h,u_y=0$) controlled by local term $H^{sr}$ (Fig.~\ref{fig2}c) yields the
 { weak localization} correction  \cite{GLK}
\be
\delta g_{\rm wl} = - \delta {\ell}/\left(2\pi^2\right);
\label{gwl}
\ee
Since $A<1$, long hops \rref{g1} never overcome weak
localization but can almost compensate it if $|A-1| \ll 1$.

In this case, it is worthwhile to evaluate  the two-loop
contribution. When included as intermediate steps into
interference contribution of Fig.~\ref{fig2}c, the long range hops simply change
the bare diffusion constant $D$. However, $\delta g_{\rm wl}$ does not
contain $D$ at all,
{\em i.e.} irreducible interference processes are important. One
of them, Fig.~\ref{fig2}e, vanishes as the path $(1)$ interferes
destructively with path $(2)$ and constructively with paths $(3),(4)$.
It is the long hop part of the Hamiltonian, Fig.~\ref{fig2}d, that
leads to the logarithmic correction to $g$.
Modified non-linear
$\sigma$-model described below yields a new RG
equation for GOE
\begin{equation}
\frac{\partial g}{\partial \ell}=\frac{A-1}{2\pi ^{2}}+\frac{A}{8\pi
  ^{4}g}
+{\cal O}(1/g^2), \ g \gg 1.
\label{a26}
\end{equation}%
yielding the RG flow of Fig.~\ref{fig1}a
and the  {\em stable} fixed line
\begin{equation}
g_{c}\left( A\right) ={A}/\left[{4\pi ^{2}\left( 1-A\right) }\right].  \label{a27}
\end{equation}%
Note, that $g_{c}\gg 1$ in the limit $1-A\ll 1$ {\em i.e.}
the two loop approximation is sufficient.
At $g<1$, \req{a27} is not applicable. However, there
is a strong reason to believe that the critical line
terminates at $g=0,\ A=0$ point. Indeed, if $g<1$, the
local part of the evolution can produce only small contributions $\delta g
\simeq \delta \ell g\ln g$. Arguments leading to the estimate
\rref{g1} remain valid. Indeed, one can
chose the exact localized eigenfunctions of the 
$H_{sr}$ instead of the wave packets and obtain
\begin{equation}
{\partial g}/{\partial \ell}={A}/({2\pi ^{2}})
+{\cal O}(g\ln g) > 0, \ g \ll 1,
\label{a270}
\end{equation}%
{\em i.e.}, $g=0$ is unstable for any $A>0$. As
$\partial g/\partial \ell <0$ for $g\gg 1$, see \req{a26}, the fixed line has
the form of Fig.~\ref{fig1}a.

Application of the magnetic field removes the time reversal symmetry,
$h,u_y\neq 0$, and thus suppresses the contribution of time reversal paths
of Fig.~\ref{fig2} c,d at distances larger than
$L_h=\sqrt{D/\omega_h}$, where $\omega_h \simeq h^2\tau+ \pi\nu u^2_y$.
The interference between path $(1)$ and paths
$(3,4)$ is also destroyed and there is no more cancelation of the
local interferences in the second loop.
We find for GUE
\begin{equation}
\frac{\partial g}{\partial \ell}=\frac{A}{2\pi ^{2}}-\frac{1}{8\pi
  ^{4}g}, \ g\gg 1,
\label{a28}
\end{equation}%
similarly to results of Ref.~\cite{et} for an
electron in random magnetic field with long-range correlations.

Equation \rref{a28} has an {\em unstable} fixed line
\begin{equation}
g_{mit}\left( A\right) =\left( 4\pi ^{2}A\right) ^{-1},  \label{a29}
\end{equation}%
corresponding to the metal-insulator transition. This result is
controllable for $A\ll 1$. The RG flow of Fig.~\ref{fig2}b
follows from \req{a29} and \req{a270} for $g<1$.

Let us discuss the relation of our results to 
earlier works. At first glance, the long jumps are
equivalent to artificial models with long-range random links known as random
band matrices, RBM \cite{bm,KM},
(the matrix elements are the Gaussian variable with $\langle
h_{ij}^2\rangle \simeq 1/|i-j|^\alpha$, $\langle h_{ij}\rangle=0$)
 The dipole hop model
\rref{dipole-hop} is different as it contains an infinite number of
the long hop loops ({\em e.g.} the product of the matrix elements
$\langle h_{12}h_{23}h_{34}h_{41}\rangle$ for non-coinciding points $1,\dots,4$
vanishes for RBM and it is finite for the dipolar interaction) and
determines all of
the interference contributions of Fig.~\ref{fig2}d.
The quantum dipole problem with long range hops was considered by
Levitov \cite{Levitov}. He started from the strongly localized states,
$g\to 0$, and replaced
the strongly localized pairs of Fig.~\ref{fig2}b by one effective dipole on
each linear scale.
{
This approach misses the logarithmic contributions
of the
multiple short range hops. We believe that in such a way one
can show the instability of $g=0$ state but can not
obtain the critical line ($g \propto \lambda$
claimed in Ref.~\cite{Levitov2} is neither
Thouless conductance nor ac-conductance. Both those quantities are
$\propto \lambda^2$, see \req{a270}).
}

{\em Supersymmetric non-linear $\sigma$ ($nl\sigma$) model}  is usually formulated \cite{book}
in terms of the $8$-component supervector. The $8$-dimensional
space where this vector resides  can be presented as a direct product of $3$
two-dimensional subspaces, $RA$, $N$ and $g$, for the retarded-advanced, Gorkov-Nambu, and the fermion-boson sectors. The matrix
structure of the Hamiltonians \rref{dipole-hop} and \rref{H}, brings up
an additional two dimensional subspace (without subscript), so the resulting
supervector $b$ is 16-component. One more supervector $a$ with the
same structure  as $b$ is
introduced to decouple the long-range hops as it is shown in \req{local}.
Those two vectors can be united in one
$32$-dimensional
supervector $\psi^T=(a^T,b^T)^{ab}$. The disorder averaging is
then performed as
\begin{subequations}
\label{a11}
\be
\begin{split}
&\langle \dots \rangle =
\int \cdots\exp \left( -L\left[ \psi \right] \right){\mathcal D}\psi,
\ \psi^\dagger\hat{\boldsymbol{\Lambda}}=\bar{\psi}=\left[\hat{\mathbb
    C}\psi\right]^T,
\\
&
\hat{\boldsymbol{\Lambda}}=\hat{\Lambda}\otimes\openone^{ab}\otimes\openone;\ \
\hat{\Lambda}=\hsigma{z}{RA}\otimes \openone^{N} \otimes \openone^{g};
\\
& \hat{\mathbb C}=\hat{C} \otimes
\openone^{ab}\otimes\openone;
\ \hat{C}=
\openone^{RA}\otimes \left(
\hsigma{-}{N}\otimes\openone^g-\hsigma{+}{N}\otimes\hsigma{z}{g} \right),
\end{split}
\raisetag{30pt}
\label{a11a}
\ee
where $\dots$ in the LHS  stands for any combination
of advanced/retarded Green functions $\hat{G}^{A,R}=\left(
\varepsilon\mp\omega/2 -\hat{H}_0-\hat{V}-\hat{H}_{lr} \mp i0 \right) ^{-1}$
 and  $\cdots$ in the RHS for the corresponding sources \cite{book},
whose form is not important for RG.
Term $L_0$ in the Lagrangian $L=L_0+L_{int}$ describes  the
property of the clean system
\be
\begin{split}
& L_0\left[ \psi \right]
 = \int d^2\r \left\{ \bar{\psi}\left[{i\lambda}\hat{\mathbb M}/2 +
  \hat{\boldsymbol{\Lambda}} 0^+\right]\psi +i\bar{\b}\hat{{\mathbb H}}_0(-i\nnabla)\b \right\};
\\
&
{\mathbb H}_0
= \left[
\left(\frac{k^2}{2m_1}-\epsilon\right)  \hat{I} -
\frac{\omega}{2}\hLambda\right] \otimes \openone
+ \left[\frac{\k^T \otimes \k}{2m_2}\right] \otimes \hat{I}
\\
&
+ h \hsigma{y}{}\otimes \hat{\Sigma}_3;\ \
 \hat{\mathbb M} = \begin{pmatrix} \nnabla^2 \hsigma{z}{}  \ &
    \nabla_\alpha  \hsigma{\alpha}{} \\ -   \nabla_\alpha  \hsigma{\alpha}{}  & \openone^{}
\end{pmatrix}^{ab}\otimes \hat{I},
\end{split}
\raisetag{1cm}
\label{a11a}
\ee
where $\hat{I} \equiv  \openone^{RA}\otimes \openone^{N} \otimes
\openone^{g}$, and $\hat{\Sigma}_3=\openone^{RA}\otimes \hsigma{z}{N} \otimes
\openone^{g}$.
The  term $L_{int}$ originates from the disorder
averaging \rref{a9}:
\be
L_{int}\left[ \psi \right] =\frac{1}{2}\sum_{i=0,x,y,z} u_i^2\int
 \left( \bar{
b}\hsigma{i}{}b \right)^2d^2{\r}.
\label{a11c}
\ee
\end{subequations}

As for the conventional $nl\sigma$-model the interaction is decoupled by introduction
the additional field $\hat{Q} \propto \b \otimes \bar{\b}$ and fixing $Q^2=\hat{I}$ in the saddle point approximation is valid as long as $\epsilon\tau \gg 1$.
For $\a=0$,
the field $\b$ for a fixed $Q$ is massive and
can be integrated out producing convergent gradient expansion in
$nl\sigma$ model. However, field $\a$
is massless and its fluctuations lead to the logarithmic corrections
discussed above. Therefore, only ``fast'' field  $\psi$ with the
momenta close to the energy shell  $k_{\pm}(\epsilon)$ can be included
into the effective theory for $Q$, whereas fields with the smaller
momenta   must be kept in the theory:
\begin{subequations}
\label{a110}
\be
\begin{split}
&\langle\dots\rangle=\int \cdots e^{ -F\left[\hat{Q} \right]-
 {\cal L}\left[\psi, \hat{Q} \right]}
{\mathcal D}\hat{Q}{\mathcal D}\psi ,\  \hat{Q}^2\!=\hat{I};
\\
&\ \hat{Q}=\hat{C}\hat{Q}^T\hat{C}^T\!;\
 \hat{Q}^\dagger=\hat{\mathbb K}\hat{Q}\hat{\mathbb K};
\ \hat{\mathbb K}=
\begin{pmatrix}\openone^{g}& 0\\0 &\hsigma{z}{g}& \end{pmatrix}^{RA} \otimes \openone^{N}.
\end{split}
\raisetag{1.7cm}
\label{a110a}
\ee
Here, $\psi$ is the 32-component {\em smooth}
on the scale of the mean free path
field satisfying the
constraints of \req{a11a};
$\hat{Q}$ is the smooth $8\times 8$ supermatrix \cite{book}.
The entries in \req{a110a} are
\be
\begin{split}
&F =\frac{\pi }{8}{\rm str}\int d^2\r \left\{
g \left(\nnabla \hat{Q}\right) ^{2} + \nu
\left[2i \omega \hat{\Lambda} \hat{Q} -  \omega_h
\left(\hat{\Sigma_3}\hat{Q}\right)^2\right]
\right\};
\\
&{\cal L}=\frac{i\lambda}{2}\int d^2\r\,
\bar{\psi}(\r)\mathbb{B}(\hat{Q}){\psi}(\r);
\\
& \hat{\mathbb B} =
\hat{\mathbb M}+ \frac{2}{\lambda}
 \begin{pmatrix}0; & 0\\
0;  & -\left(\epsilon \hat{I} +\frac{i\hat{Q}(\r)}{2\tau} \right)
\otimes \openone
+ { h}
\hsigma{y}{}
\otimes \hat{\Sigma}_3
\end{pmatrix}^{ab},
\end{split}
\raisetag{1.8cm}
\label{a110b}
\ee
\end{subequations}
where the bare conductance is given by
\[
g=\sum_{\pm} \int \frac{d^2{\k}}{\left( 2\pi \right) ^{2}}
\frac{\tau}{2}\frac{\partial E_\pm}{\partial k_\alpha}\frac{\partial E_\pm}{\partial k_\alpha}\delta \left[ \epsilon - E_\pm (\k)\right] .
\]

Note that, in contrast to non-local $RBM$ models \cite{bm}, the theory \rref{a110} is renormalizable. Indeed, $F[\hat{Q}]$ of
\req{a110a} includes only relevant terms allowed by
$\hat{Q}^2=\hat{I}$. The same applies to the Lagrangian $\mathcal{L}$ as the
natural dimensions of fields $[\nabla \a] = [\b]=1$ while
$\a$ itself can not enter: $\mathcal{L}$ is invariant with respect to constant shift of $\a$.

We checked that  the fluctuations of $Q$-matrix do not change the
Lagrangian  $\mathcal{L}$ and, thus, the coefficient $A$ in \reqs{a26}-\rref{a28}.
This relates to the fact
that the averaged density of states cannot
have corrections from mixing of retarded and advanced sectors
described by $Q$ matrix.

In order to sum up leading logarithmic divergences, we performed two-loop RG analysis of theory \rref{a110b}, see Ref.~\cite{preparation} for details.
We represent $\hat{Q}=\hat{\bar{V}}\left[(1+i\hat{P})/(1-i\hat{P})\right]V$
and $\psi=\psi_<+\hat{V}\psi_>$, where $\psi_<,\hat{V}$
are slow variables and the
other are fast. We integrated out the fast variables in the second
loop approximation to obtain the
effective free energy for slow ones.
Varying the result of the integration with
respect to a gauge invariant cutoff
 we obtain \reqs{a26} and \rref{a28}.

{\em In conclusion,}
we performed the scaling analysis of the localization problem of
2D quantum dipoles, see Fig.~\ref{fig1}.
For
T-invariant systems, the Thouless conductance  tends to a
finite  value. Breaking the T-invariance leads to the
transition between the antilocalization and a critical behavior.
To describe those phenomena
we constructed a novel version
of the non-linear $\sigma$-model, \req{a110},
and derived the second loop RG equations. This model  also allows
studies  \cite{preparation} of the multifractal properties of the wave-function
 using methods of Ref.~\cite{book}.

We are grateful to L.I. Glazman for reading the manuscript and useful remarks.
Support by  US DOE contract No. DE-AC02-06CH11357 (I.L.A. and B.L.A.),
NSF-CCF Award 1017244 (B.L.A),
and Transregio 12 of DFG (K.B.E.) is  acknowledged.

\end{document}